\documentclass{article}
\usepackage{amssymb}
\usepackage{amsmath}

\newcommand{\beq}{\begin{equation}}
\newcommand{\eeq}{\end{equation}}
\newcommand{\bea}{\begin{eqnarray}}
\newcommand{\eea}{\end{eqnarray}}
\newcommand{\rref}[1]{{(\ref{#1})}}
\newcommand{\dsl}[1]{{\displaystyle{#1}}}
\newcommand{\la}{{\lambda}}
\newtheorem{theorem}{Theorem}
\newtheorem{prop}[theorem]{Proposition}

\newtheorem{lem}[theorem]{Lemma}

\newtheorem{remark}[theorem]{Remark}

\def\endpf{\begin{flushright}$\square$\end{flushright}}

\begin{document}

\title{The Sato Grassmannian and the CH hierarchy}
\author{Gregorio FALQUI $^\bullet$, Giovanni ORTENZI $^\odot$}
\date{}
\maketitle
\noindent
$^\bullet $ Dipartimento di Matematica e Applicazioni,
Universit\`a di Milano-Bicocca
Via R. Cozzi 53 -- Ed. U5, I-20126 Milano, Italy\\
E-mail: gregorio.falqui@unimib.it\\[10pt]
$^\odot$
Dipartimento di Fisica Nucleare e Teorica,
Universit\`a di Pavia,
Via A. Bassi, 6  27100 Pavia, Italy, and
I.N.F.N. Sezione di Pavia\\
E-mail:giovanni.ortenzi@unimib.it
\begin{abstract}
\noindent
{We discuss how the Camassa-Holm hierarchy can be framed within the
geometry of the Sato Grassmannian.}
\end{abstract}

\section{Introduction}
In this paper we study some specific aspects of the Camassa Holm
hierarchy. Since it appearance in the literature, it has been
recognized that the CH equation possesses specific features, (e.g.,
peakon solutions, the appearance of third order Abelian
differentials in finite gap solutions,...) that other more
"classical" soliton hierarchies (KdV, Boussinesq, NLS) do not
exhibit. Among these, especially in view of the Dubrovin--Zhang
classification scheme \cite{DubZh}, the non-existence of a
formulation via a $\tau$ function is, from our point of view, of
particular interest. The Sato theory of the $\tau$ function,
basically views it as a section of the (dual) determinant bundle
over the so--called Sato (or Universal) Grassmannian (UG), and
allows to associate such a structure to any hierarchy of
evolutionary PDEs that can be represented as linear flows on this
Grassmannian. Thus, it seems important to analyze whether (and
which) flows of the CH hierarchy can be realized as linear flows in
the Sato Grassmannian.

The main aim of this paper is to discuss this problem, in the
framework of a set up, introduced in \cite{FMP, CFMPBrasil},
relating the (bi)--Hamiltonian structures  of soliton hierarchies of
KdV type to the Sato Grassmannian.\\
In \cite{KM} it was shown that the bi--Hamiltonian structures of CH
and KdV equations (as well of the Harry--Dym equation) are related,
being geodesic motions on the Virasoro group with respect to
different metrics. Actually, the relation with the evolution on the
Sato Grassmannian has been studied for KdV and the HD hierarchies,
showing that they are related to linear flows in the big cell of UG.
In this paper we try to complete this picture showing that the CH
hierarchy too is related to the big cell of the Sato Grassmannian by
means of its local (also called negative) flows.\\
One of the basic differences among this representation of the three
hierarchies is given by the relation between the local flows and the
``time'' of the hierarchy related to the conservation of the linear
momentum.\\
This will show up, in the present paper, as the
realization of the CH local hierarchy in a constrained subspace of
the big cell.
The path leading us to this result is the analysis of the evolution
of the Noether currents associated with the bi--Hamiltonian
recurrence relation of the local hierarchy. We will argue as, on
more general grounds, they are associated with a two-field (albeit
somehow trivial) bi--Hamiltonian extension of the CH local
hierarchy.\\
The ordinary CH bi--Hamiltonian hierarchy is recovered -- together
with the non local part including the "true" CH equation -- by Dirac
restricting this two-field hierarchy to a specific submanifold,
namely those selected by these Noether currents satisfying a
specific constraint.\\
Thus the CH equation is realized, in the picture herewith presented,
as an additional commuting flow of an infinite system of linear
flows on the Sato Grassmannian.

The full interpretation of the whole nonlocal hierarchy to this Sato
Grassmannian approach, as well as the problem of how far this
picture could be useful to explain and understand the non--existence
of the $\tau$ function for CH is still under consideration.
\section{The geometry of the CH hierarchy .... } It is well
known\cite{CamHol, FF81} that the CH equation\footnote{We have
herewith chosen unusual normalizations because this somewhat
simplifies some of the formul\ae\ we are interested in.}
$$
4 v_t- v_{xxt}=24v_xv-4v_{xx}v_x-2vv_{xxx}
$$
is a bi--Hamiltonian evolutionary PDE on
$C^{\infty}(S^1,\mathbb{R})$ w.r.t. the Poisson pencil
$$
P_{\lambda}= (4 \partial_x -\partial_x^3)+ \lambda (2m\partial_x + 2\partial_x m)
 \qquad \lambda \in \mathbb{R}
$$
where $m=4v- v_{xx}$.
\\
The densities of the conserved laws of the hierarchy can be obtained by
recursively solving
\begin{equation}
\label{RicCH}
h_x+h^2=m z^2+1, \qquad z=\sqrt{\lambda}
\end{equation}
where $h$ is the generating function of
the densities of the Casimir of $P_\lambda$ \cite{CLOP,C,CmK,Len,R}.\\
This Riccati equation admits two different solutions
\begin{eqnarray*}
&&h=h_{-1}z+h_0+\frac{h_1}{z }+\frac{h_2}{z^2 }+\dots \\
&&k=k_0+k_{-1}z+k_{-2}z^2+k_{-3}z^3+\dots \quad .
\end{eqnarray*}
The two families of coefficients
$\{ h_i \}_{i \geq -1}$ and $\{k_i\}_{i \leq
  0}$ give, by means of the Lenard recursion,
all the f CH hierarchy. In particular, the $h_i$'s are the densities of the
negative (or local) CH hierarchy, and can be algebraically found from
(\ref{RicCH}), while the $k_j$'s are the densities of the positive (or
``non--local'')   CH hierarchy, whose first two members are, respectively
$x$-translation and the CH equation itself.
\\
The first flow of the local hierarchy is
\begin{equation}
\label{locCH}
\frac{\partial}{\partial t_3} m
=(4\partial_x -  \partial_x^3) \frac{1}{2\sqrt{m}}.
\end{equation}
The key ingredient used in \cite{FMP} to relate the Hamiltonian
structure of Soliton hierarchies of KdV type to evolutions on the
Sato Universal Grassmannian manifold is given by the Noether
currents.

In particular, it has been shown in \cite{CLOP} that the Noether currents
associated with the local CH hierarchy are characterized, in the space of
formal Laurent series in the parameter $z$ by the following two properties:
\begin{enumerate}
\item Their asymptotic behavior is given by
\begin{equation}
\label{locJCH}
J^{(s)} = z^{s}+O(z) , \quad s\geq 2
\end{equation}
\item
They belong to  the span
\begin{equation}
\label{FdB}
\langle (\partial_x + h)^n z^2 \rangle_{n \geq 0}
 \end{equation}
of the {\it Fa\`a di Bruno} monomials associated with the generating
function $h$, which solves \rref{RicCH} with asymptotic condition
$h(z)=h_1z+h_0+\dsl{\frac{h_1}{z}}+\cdots$, with coefficients on
$C^{\infty}(S^1,\mathbb{R})$.
\end{enumerate}
The connection between the currents $J^{(s)}$ and the generating
function $h$ is given by the fact that, along the $s$-th time of the
local CH hierarchy, they evolve as
\begin{equation}\label{eq:cl}
\partial_s h=\partial_x J^{(s)} \qquad \mathrm{where} \quad \partial_s=\frac{\partial}{\partial t_s}.
\end{equation}
The asymptotic behavior of the local Noether currents and the
presence of a ``generator'' $h$ suggest, in analogy with what
happens in the KdV case, that they can  be associated with linear
evolutions on the Sato Grassmannian.
\section{.... and the Sato Grassmannian}
In this section we shall look at the problem starting from a
slightly different
perspective. \\
Let us consider the space $J_+$ given by the span
on $C^{\infty}(S^1,\mathbb{R})$  of the family
$$
J^{(i)}=z^i + J^i_{-1} z +J^i_{0}  +J^i_{1} z^{-1} + \dots \qquad i \geq 2
$$
in
the space $J$ of Laurent series (with at most a pole singularity at
$z=\infty$). $J$ admits a direct splitting as
\begin{equation}
\label{split}
J=J_+ \oplus J_-, \quad \mathrm{where} \quad J_- :=\langle z^i\rangle_{i \leq 1}.
\end{equation}
Therefore  the collection $\{J^{(i)}\}_{i \geq 2}$ defines a point
of the big cell $\mathcal{B}$ of the Sato Grassmannian translated by
$z^2$ w.r.t. the standard Sato representation \cite{SW}.\\
On this space we can define an infinite family of flows setting
\begin{equation}
\label{CS+2}
(\partial_s + J^{(s)}) J_+ \subset J_+  \qquad s \geq 2
\end{equation}
that, more explicitly, can be written as
\begin{equation}
\label{CS+2comp}
(\partial_s+J^{(s)})J^{(r)}= J^{(s+r)} +\sum_{i=-1}^{r-2}J^s_iJ^{r-i}
                          +\sum_{i=-1}^{s-2}J^r_iJ^{s-i}
                          +J^{r}_{-1} J^{s}_{-1} J^{(2)} \ .
\end{equation}
\begin{prop}
\label{propCS+2int}
The flows (\ref{CS+2}) commute.
\end{prop}
{\bf Proof } We have to show that $[\partial_s, \partial_r]J_+ = 0$, i.e.
\begin{equation}
\label{commCS+2}
[\partial_s, \partial_r]J^{(n)} = 0, \qquad \forall s,r,n \geq 2 \ .
\end{equation}
Thanks to (\ref{CS+2}) it holds
the symmetry $\partial_s J^{(r)}= \partial_r J^{(s)}$
and then the equation (\ref{commCS+2}) can be written as
\begin{equation}
\label{commplusCS+2}
[\partial_s, \partial_r]J^{(n)}=[\partial_s+J^{(s)}, \partial_r+J^{(r)}]J^{(n)}.
\end{equation}
From the explicit form of the currents it holds
$$
[\partial_s, \partial_r]J^{(n)} \in J_-,
$$
but from (\ref{CS+2})
$$
[\partial_s+J^{(s)}, \partial_r+J^{(r)}]J^{(n)} \in J_+.
$$
\endpf
\begin{prop}\label{currCH}
The local currents of CH satisfy (\ref{CS+2}).
\end{prop}
{\bf Proof } The currents (\ref{locJCH}) are elements of $J_+$.
Moreover from the property (\ref{FdB}) follows that  every element
of $J_+$  can be written as  $J^{(i)}_{lCH}=\sum_k c^i_k
(\partial_x+h)^k z^2$. Using  this expansion \rref{eq:cl} we see
that
\[
\begin{split}
&(\partial_s+J^{(s)})\sum_{k=0}^r c^r_k (\partial_x+h)^k z^2 =
\sum_{k=0}^r (\partial_s c^r_k)(\partial_x+h)^k z^2 +\sum_{k=0}^r
c_k^r (\partial_s+J^{(s)})(\partial_x+h)^k z^2 \\ &= \sum_{k=0}^r
(\partial_s c^r_k)(\partial_x+h)^k z^2+\sum_{k=0}^r (\partial_x+h)^k
z^2\,J^{(s)} \subset J_+ \oplus z^2 J_+ \ .\end{split}
\]
In \cite{CLOP} it is shown that, for the local currents of CH, $ z^2 J_+ \subset  J_+ $ and then they satisfy (\ref{CS+2}).
\endpf
Therefore, taking into account the results of \cite{FMP} we can
conclude that the local (negative) flows of CH hierarchy are given,
by means of the construction outlined above, linear flows on
the big cell $\mathcal{B}$ of the Grassmannian. \\
{\bf Remark}. The basic issue to recover a hierarchy of 1+1
dimensional PDEs from a dynamical system of the form \rref{CS+2} is
to specify (or define) the ``physical'' space variable $x$.

For instance, in the ordinary KP-KdV case, $x$ can be, as it is well
known, identified with the first "time" of the hierarchy. As it was
shown in \cite{CFMP6}, fractional KdV hierarchies can be obtained
identifying $x$ with a different time $t_s$ of a system similar to
\rref{CS+2}. Actually, in our case, $x$ is not contained in the
dynamical system, and thus should be added by means of the
introduction of another current $h=h_{-1}z+h_0+\frac{h_1}{z}+\dots$.
In turn, this additional current has to be related with the action
of $x$-translation on the currents $J^{(s)}$ of the Grassmannian.

The most natural way to add this new current is to consider the
enlargement of the system (\ref{CS+2}) to
\begin{equation}
\label{nnoint}
(\partial_s + J^{(s)}) J_+ \subset J_+,  \qquad
(\partial_s+J^{(s)})h \in J_+ \>\> (s \geq 2), \qquad
(\partial_x+h)J_+ \subset J_+ ,
\end{equation}
which explicitly is given, in addition to Eqn.s \rref{CS+2comp}, by
\begin{equation}
\label{noint}
\begin{split}
(\partial_x+h)J^{(s)}&= \sum_{i=-1}^{s-2} h_{i} J^{(s-i)}+  h_{-1}
J^s_{-1} J^{(2)}
\qquad s \geq 2 \\
(\partial_s+J^{(s)})h&= \sum_{i=-1}^{s-2} h_{i} J^{(s-i)}+ h_{-1}
J^s_{-1} J^{(2)} \qquad s \geq 2.
\end{split}
\end{equation}
However, these flows are not in general commuting, so that further
conditions have to be imposed. It is outside the size of this paper
to discuss this problem in full generality; we simply remark the
restriction to the subspace of the translated big cell defined by
\begin{equation}
\label{ourcon}
 J^{(2)}=z^2 \qquad and \qquad z^2 J_{+} \subset J_{+} \ .
\end{equation}
is a consistent one\footnote{Another consistent solution to this
problem is given by requiring that $ (\partial_x+h)h \in J_+ $. The
resulting system of commuting PDEs leads to a $2+1$ dimensional
extension of the HD hierarchy \cite{KO,PSZ}.}.

The following Lemma helps clarifying the meaning of the
constraint\rref{ourcon}:
\begin{lem}
\label{reclem} For any choice of $J^{(2)}$, the currents $J^{(i)}$
satisfying (\ref{noint}) are elements of $F=sp\langle (\partial_x
+h)^n J^{(2)} \rangle_{n \geq 0}$.
\end{lem}
{\bf Proof } Expanding the relation (\ref{noint}) it follows that
\begin{equation}
\label{rec}
J^{(s+1)}=\frac{1}{h_{-1}} (\partial_x+h)J^{(s)} - \sum_{i=-1}^{s-2} \frac{h_{i}}{h_{-1}} J^{(s-i)}+  J^s_{-1} J^{(2)}.
\end{equation}
Since $(\partial_x+h)F \subset F$ and $J^{(2)} \in F$, then one can
write recursively all the currents using elements of $F$.
\endpf
In the light of this proposition, we can rephrase the first of
equations \rref{ourcon} saying that we consider only the case
$J^{(2)}=z^2$. The study of more general choices of the current
$J^{(2)}$ is under consideration.

The basic reason for this choice of ours is that the space $J_{+}$
defined by \rref{ourcon} contains the currents of the CH hierarchy
(see Proposition \ref{currCH}). Moreover, it turns out that $J_{+}$
is parameterized by three fields, namely $h_{-1}$, $h_{0}$,  and
$h_{1}$. This can be seen as follows. Since $z^2\,J_{+}\subset
J_{+}$ and $J^{(2)}=z^2$, we get that $J^{(4)}=z^4$. The recursion
relations (\ref{rec}) allow us to write all the currents, and
namely $J^{(4)}$, as differential polynomials in the components $h_k$
of the formal Laurent series $h$. Thus we arrive at
\begin{equation}
\label{constrJ4}
{\frac {z^2}{{h_{{-1}}}^{2}}} (h_x+h^2)
- z^2 \left( {\frac {  h_{{-1}_x} }{{h_{{-1}}}^{3}}}+{\frac {2 h_{{0}}}{h_{{-1}}}^2} \right) h
- z^2 \left( {\frac { h_{{0}_x}}{{h_{{-1}}}^{2}}}-{\frac {{h_{{0}}}^{2}}{{h_{{-1}}}^{2}}}
+{\frac {2h_{{1}}}{h_{{-1}}}}-{\frac {h_{{0}} \left( h_{{-1}_x} \right) }{{h_{{-1}}}^{3}}} \right) =z^4.
\end{equation}
It is straightforward to check that this relation enables one to
recover $h_2,h_3,\ldots$ as differential polynomials in
$h_{-1},h_0,h_1$. So the system \rref{noint}, determines a hierarchy
of 1+1 evolutionary PDEs in the three fields (dependent variables)
$h_{-1},h_0,h_1$. For instance, the first non trivial flow is
\cite{OC}:
\begin{equation}\label{eqt3}
\begin{split}
\partial_3 h_{-1}=&-{\frac {  h_{{-1}_x}  h_{{1}}}{{h_{{-1}}}^{2}}}+{\frac {  h_{{1}_x}  }{h_{{-1}}}} \\
 \partial_3 h_{0}=& \frac{3}{2}{\frac {  h_{{1}_x}  h_{{-1}_x}  }{{h_{{-1}}}^{3}}}
-\frac{3}{2}{\frac {h_{{1}} ({h_{{-1}_x}})^{2}}{{h_{{-1}}}^{4}}}
-\frac{1}{2}{\frac {h_{{1}_{xx}}  }{{h_{{-1}}}^{2}}}+\frac{1}{2}{\frac { h_{{-1}_{xx}} h_{{1}}}{{h_{{-1}}}^{3}}}\\
 \partial_3 h_{1}=&-\frac{3}{2}{\frac { h_{{-1}_x}  h_{{1}_{xx}}  }{{h_{{-1}}}^{4}}}
+\frac{5}{2}{\frac { h_{{-1}_x}      h_{{-1}_{xx}} h_{{1}}}{{h_{{-1}}}^{5}}}
+{\frac {15}{4}}\,{\frac { \left(  h_{{-1}_x}  \right) ^{2}  h_{{1}_x}  }{{h_{{-1}}}^{5}}}
-{\frac {15}{4}}\,{\frac {h_{{1}} \left(   h_{{-1}_x}  \right) ^{3}}{{h_{{-1}}}^{6}}}\\
&+{\frac {{h_{{1}}}^{2}  h_{{-1}_x} }{{h_{{-1}}}^{3}}}
+\frac{1}{4}{\frac {  h_{{1}_{xxx}} }{{h_{{-1}}}^{3}}}
-\frac{1}{4}{\frac { h_{{-1}_{xxx}} h_{{1}}}{{h_{{-1}}}^{4}}}
-{\frac {  h_{{1}_{x}} h_{{-1}_{xx}} }{{h_{{-1}}}^{4}}}
-{\frac { h_{{1}_x}  h_{{1}}}{{h_{{-1}}}^{2}}} \qquad .
\end{split}
\end{equation}
We notice that the field $h_0$ does not affect the dynamics . Actually,
this is true for all the times of the hierarchy we are considering.
This is a consequence of the fact that no currents depends on $h_0$,
as one can see by recursion using (\ref{rec}).

Therefore the constraint given by (\ref{constrJ4}) do not depend on
$h_0$ as well, and so we can limit ourselves to the study of the
system in the two dependent variables $h_{-1},h_1$.

We shall prove that this system is bi--Hamiltonian and admits an
iterable Casimir, that is, a Casimir of the pencil that generates,
via the Lenard recursion relations, the commuting flows. Our proof
will be done in a sequence of steps as follows.

First we notice that, if we perform the change of variables
$h_{-1}=\alpha$ and $h_1=\frac{\gamma}{\alpha}$ the first and third of equations
\rref{eqt3}  become:
\begin{eqnarray}
\label{alphagamma}
&&\partial_3 \alpha=\left( \frac{\gamma}{\alpha^2}\right)_x \nonumber\\
&&\partial_3 \gamma=
\frac{\alpha}{4} \left( \frac{1}{\alpha}\left( \frac{1}{\alpha}\left( \frac{\gamma}{\alpha^2}\right)_x\right)_x\right)_x .
\end{eqnarray}
From the general theory, and namely from the representation
\rref{eq:cl} of the PDEs, we see that this system has an infinite
sequence of conserved quantities, whose densities are given by the
coefficients of the formal Laurent series (\ref{constrJ4}) with
$h_0=0$, i.e.:
\begin{equation}
\label{constrJ4h0=0}
\frac{1}{\alpha^2}(h_x+h^2)-\frac{\alpha_x}{\alpha^3}h-\frac{2\gamma}{\alpha^2}=z^2.
\end{equation}
It is worthwhile to remark again that this equation determines all
the coefficients $h_i, i\ge 0$ as differential polynomials in
$\alpha,\gamma$. For instance we have, apart form the obvious
relations $h_{-1}=-\alpha, \> h_{1}=-\gamma/\alpha$, the expressions
\begin{equation}\label{ha3}
\begin{split}
h_2&=\left(\frac{\gamma}{2\alpha^2} \right)_x,\quad
h_3=\dsl{\frac{\gamma^2}{2\alpha^3}}-\left(\dsl{\frac1{\alpha}\left(\frac{\gamma}{\alpha^2}\right)_x}\right)_x,\quad
h_4=\ total \ derivative,\\
h_5 &= \frac{\gamma^3}{2\alpha^5} 
-\frac{1}{12}\frac{\gamma^2\alpha_{xx}}{\alpha^6}
+\frac{1}{8}\frac{\gamma\gamma_{xx}}{\alpha^5}
-\frac{7}{24}\frac{\gamma\gamma_x\alpha_x}{\alpha^6}
+\ total \ derivative, \dots 
\end{split}
\end{equation}
and so on and so forth.\\
The motivation for the change of variables, as well as further hints
for our program come from considering of the dispersionless limit of
(\ref{alphagamma}), that is,
\begin{eqnarray}
\label{alphagammadisp}
&&\partial_3 \alpha=\left( \frac{\gamma}{\alpha^2}\right)_x \nonumber\\
&&\partial_3 \gamma=0.
\end{eqnarray}
This equation is bi--Hamiltonian w.r.t. to the Poisson tensors
\begin{equation}\label{Pagdisp}
P_0^{disp}= \left( \begin{array}{ccc} 0 && \partial_x \alpha\\
&&\\\alpha\partial_x&&
\gamma\partial_x+\partial_x\gamma
\end{array} \right)
 \qquad
P_1^{disp}=\left( \begin{array}{ccc}
\partial_x &&
0
\\ &&\\
0 && 0
\end{array} \right)
\end{equation}
with Hamiltonian densities $ \ h_{3}=\gamma^2/2\alpha^3,
h_{1}=-\gamma/\alpha$. This property suggests that the full
dispersive hierarchy can be obtained by suitably deforming the
pencil of Poisson tensors (\ref{Pagdisp}).

As a first step in this direction, one notices that the flow
\rref{alphagamma} can be obtained in a "Hamiltonian" way, via the
action of the antisymmetric tensors
\begin{equation}\label{Pag}
P_0=
\left( \begin{array}{ccc} 0&&\partial_x \alpha\\
&&\\\alpha\partial_x&&
\gamma\partial_x+\partial_x\gamma+\frac{\alpha}{4}\partial_xT_\alpha^2\alpha
\end{array} \right),
\qquad P_1=\left( \begin{array}{ccc}
\partial_x &&
\frac14\partial_xT_\alpha^2\alpha
\\ &&\\
\frac{\alpha}{4}\partial_xT_\alpha^2&&
\frac{\alpha}{16}\partial_x T_\alpha^4\alpha
\end{array} \right),
\end{equation}
where $T_\alpha$ is the operator\footnote{Operator composition is
here and in the following, understood.}
$\dsl{\frac1{\alpha}\partial_x}$, as
\[
\left( \begin{array}{c}
\partial_3 \alpha\\
\partial_3 \gamma\end{array}\right)=P_0 d\int h_3\, dx =P_1\int
h_1\,dx\>,
\]
where $h_1$ and $h_3$ are the densities \rref{ha3}. Furthermore, a
direct computation shows that $h_1$ is the density of a Casimir of
$P_0$. Actually, our use of this terminology is justified by the
following proposition, whose proof, that can be directly obtained via a
straightforward albeit tedious computation, will be apparent from the sequel.
\begin{prop} The tensors \rref{Pag} are a pair of compatible Poisson
tensors.
\end{prop}
To push our analysis further, the following observation is
important. We notice that the member $P_1$ of the pair \rref{Pag} is
greatly degenerate. Indeed one sees that vector fields
$(\dot\alpha,\dot\gamma)$ belong to its image if and only if the
relation
\begin{equation}
  \label{eq:triv}
\dot\gamma=\dsl{\frac{\alpha}{4}\partial_x\frac1{\alpha}\partial_x\frac1{\alpha}\,\dot\alpha}(=
\dsl{\frac{\alpha}{4}\big(T_\alpha^\dagger\big)^2\,\,\dot\alpha}).
\end{equation}
This entail that the system (\ref{alphagamma}),
as well as {\em any} bi--Hamiltonian vector field associated with the pair
\rref{Pag} admits as an invariant submanifold
the one defined by
\begin{equation}\label{eq:inv-sub}
\gamma-\frac{1}{4}\partial_x^2 \ln \alpha+\frac{1}{8}(\partial_x\ln
\alpha)^2\left(\equiv
\gamma-\frac18(\alpha(T_\alpha-T^\dagger_\alpha)T_\alpha(\alpha)\right)=const.
\end{equation}
This fact (together with the particularly simple dependence on
$\gamma$ of the relation \rref{eq:inv-sub}) prompts us to consider
the dependent variable $u=y-\frac{1}{4}\partial_x^2 \ln
\alpha+\frac{1}{8} (\partial_x\ln \alpha)^2$. In the coordinates
$(\alpha,u)$ the tensors of \rref{Pag} become
\begin{equation}
  \label{nPag}
  P_0=
\left( \begin{array}{ccc} 0&&\partial_x \alpha\\
&&\\\alpha\partial_x&& u\partial_x+\partial_x u -\dsl{\frac14}
\partial_x^3.
\end{array} \right),
\qquad P_1=\left( \begin{array}{ccc}
\partial_x && 0
\\ &&\\ 0&& 0\end{array}\right).
\end{equation}
The fact that the antisymmetric tensors we are considering indeed
make up a Poisson pair is now apparent from the theory of affine
Poisson structures on duals of Lie algebras. This new form of the
pencil will also allow us to state that the hierarchy of commuting
vector fields starting with \rref{alphagamma} is indeed a
bi--Hamiltonian hierarchy.

According to the Gel'fand--Zakharevich bi--Hamiltonian scheme, we
look for a Casimir of the pencil \rref{nPag}. This amounts to
finding an exact one-form $\Omega(\lambda)=(X(\lambda),Y(\lambda))$
that satisfies the equation
\[
(P_1-\lambda P_0)\Omega=0,\quad \text{with asymptotics }
\Omega(\lambda)=\Omega_0+\frac{\Omega_1}{\lambda}+\cdots\>,
\]
whose first element is the differential of the Casimir of $P_0$ (in
particular, with obvious meaning of the notation,
$Y_0\simeq\dsl{\frac1{\alpha}}$). So we can trade the above
equation for the system
\begin{equation}\label{eqR}
    X(\la)=\la\alpha\, Y(\la); \qquad
    \la\alpha^2Y(\la)^2+2uY(\la)^2-\frac12Y_{xx}(\la)Y(\la)+\frac14(Y_x(\la))^2=\la.
\end{equation}
In turn, the second of these equations is equivalent to the
following system
\begin{equation}\label{eqR2}
    h_x+h^2=\la\alpha^2+2\,u, \qquad
    h=\frac{z}{Y(\la)}+\frac12\frac{Y_x(\la)}{Y(\la)},
\end{equation}
where $z^2=\la,$ and $h=h_{-1}z+h_0+\dsl{\frac{h_1}{z}}+\cdots$.
 It can be easily shown that the series h(z) solving the first
of these equations is, in the sense of formal Laurent series, indeed
the potentials of the one-form $\Omega(\la)$. Also, the coefficients
$h_i$ can be algebraically computed in a recursive way.

The comparison of this Riccati equation  with the Riccati equation
associated with the local CH hierarchy suggests a further minor
coordinate change, namely to set $m=\alpha^2.$ Indeed in the
coordinates $(m,u)$ the Poisson pencil $P_1-\lambda P_0$ is
(\ref{nPag})
\begin{equation}
\label{Pmu} \left( \begin{array}{ccc} 2
(\partial_x m + m \partial_x) && 0\\ &&\\
0&&0\end{array}\right)-\la\left(\begin{array}{ccc} 0
&& \partial_x m + m \partial_x   \\ &&\\
\partial_x m + m \partial_x &&
-\dsl{\frac{1}{4}} \partial_x^3 +\partial_x u + u
\partial_x
\end{array} \right),
\end{equation}
and the corresponding Riccati equation is
\begin{equation}
\label{Ricmu} h_x+h^2=2u + m z^2, \qquad z=\sqrt{\lambda}.
\end{equation}
The vector field \rref{alphagamma} becomes simply
\begin{equation}\label{equ}
\partial_3 m = ( 2u \partial_x +  2\partial_x u  - \frac{1}{2} \partial_x^3) \frac{1}{\sqrt{m}} \\
\partial_3 u=0 \>,
\end{equation}
Summing up,the search for a Casimir of the pencil \rref{Pmu} is
reduced to the problem of solving - in the space of formal Laurent
series - the Riccati equation for
$h(z)=h_{-1}z+\dsl{\sum_{i=1}^\infty\frac{h_i}{z^i}}$. This problem
can be iteratively solved, and is equivalent, up to the total
derivative $h_0$, to (\ref{constrJ4h0=0})  written in the $u,m$
variables.\\
{\bf Remarks}. \\
{\bf 1)} On $u=\dsl{\frac{1}{2}}$ the first of the equations
\rref{equ} becomes the first nontrivial local CH flow (\ref{locCH}).\\
{\bf 2)} In the coordinates $(m,u)$ (as well as in the coordinates
$(\alpha,u)$), all vector fields of this hierarchy are somewhat
trivial, since they read
\begin{equation}
  \label{eq:ht}
  \partial_{t_i}m=\partial_x(F_i(m,u)),\qquad \partial_{t_i}u=0.
\end{equation}
This fact can be, in a sense, understood also in the framework of the theory
of reciprocal transformations. For instance, transforming the system
\rref{alphagamma} under the reciprocal transformation induced by its first
element (seen as a conservation law) yield the triangular system
\begin{eqnarray*}
&&\partial_3 U= \frac{1}{2}(UV)_z \\
&& \partial_3 V= \frac{1}{4}(V_{zzz}+6VV_z)
\end{eqnarray*}
where $dx= U dz +\frac{1}{2}UV dt_3$, $U=\frac{1}{\alpha}$, and $V=
\frac{-2 \gamma}{\alpha^2}$. To fully examine these equations in the
light of the theory of reciprocal transformations, however, is
outside the aim of the present paper \cite{FHO}.\\
{\bf 3)} As a final check of the bi--Hamiltonian analysis we
performed, we notice the following We exchange the role of the
Poisson tensors $P_0$ and $P_1$ and consider the Casimir function
$K=\int (u+m) dx$ of $P_1$. Clearly enough, the vector field $P_0
dK$ is just $x$-translation. This Casimir does not give rise to a
new Lenard sequence, since $P_0dK_0$ does not lie in the image of
$P_1$. However from the fact that $x$=translation is the image under
$P_0$ of a Casimir of $P_1$ confirms that it commutes with all the
vector field of the hierarchy, as it should be.

\section{Back to the CH hierarchy: its bi--Hamiltonian structure and its Lax representation}
As we have seen, the bi--Hamiltonian geometry of the manifold we are
considering is particularly simple: indeed, it is stratified by the
submanifolds given by $u=\kappa$ for some constant $\kappa$, and
these submanifolds are left invariant by all vector fields that are
Hamiltonian w.r.t $P_1$, and thus by all bi--Hamiltonian vector
fields. Also, on the invariant submanifold $u=\dsl{\frac12}$ we have
that the first flow of our hierarchy coincides with the first local
CH flow, and the Riccati equation \rref{Ricmu} reduces to the
Riccati equation associated with the CH hierarchy \rref{RicCH}.

These facts suggest the opportunity to consider the Dirac reduction
of the pencil \rref{Pmu}.
\begin{prop}
The Dirac reduction of (\ref{Pmu}) on the constraint $u=\kappa$
gives a Poisson pencil for the Camassa Holm. The hierarchy restricts
to this submanifold as a bi--Hamiltonian hierarchy.
\end{prop}
{\bf Proof}. To prove the assertion, we find it more convenient to
use the notation of Poisson brackets rather than that of Poisson
tensors. According with Dirac's theory, the reduction on $u=const$
of the Poisson brackets associated with our pencil is given by
\begin{eqnarray*}
\{m(x),m(y)\}^D_0&:=&\{m(x),m(y)\}|_{u=\kappa}\\
                &&-\int dw \int dz \{m(x),u(w)\} (\{u(w),u(z)\})^{-1} \{u(z),m(y)\} |_{u=\kappa}
\end{eqnarray*}
where $\{u^i(x),u^j(y)\}_0:=
\int dz \frac{\delta u^i(x) }{\delta u^k(x)  } (P_\lambda)^{kl} \frac{\delta u^j(y) }{\delta u^l(x)} $.\\
A simple computation shows that
$$
P_{\lambda}^D|_{u=\kappa}= 2  (\partial_x m + m \partial_x) -
\lambda (\partial_x m + m \partial_x) \left(2\ \kappa\ \partial_x
-\frac{1}{4}
\partial_x^3\right)^{-1}(\partial_x m + m \partial_x).
$$
It is easy to recognize in the above formula (one of) the Poisson
pencils of the CH hierarchy, namely the one given by the standard
Lie Poisson tensor and the first nonlocal tensor with the suitable
choice $\kappa=\frac{1}{2}$
\footnote{Indeed, $\kappa$ can be rescaled to $\frac{1}{2}$ without loss of generality. For $\kappa=0$, 
we get a Poisson pencil of HD.}.
The Dirac reduction of the Poisson
structure (\ref{Pmu}) generates exactly the local part of the CH
hierarchy. This follows from the fact that the Dirac deformation of
the Poisson bracket associated with $P_0$ is achieved by means of
Casimir functions of the other brackets. This entails that Lenard
relations $P_0 dH=P_1 dK$ hold also for the corresponding Dirac
reductions. On the manifold $u=\kappa$ (e.g., $u=\frac12$) we can
recover the standard nonlocal part of CH hierarchy using the
solution of (\ref{Ricmu}) whose asymptotic behavior is $1+O(z)$ as
in \cite{CLOP}, via the usual CH substitution $m=4v-v_{xx}$. In this
picture, the flows of the positive CH hierarchy (and so, the CH
equation as well) play the role of ``additional'' (commuting) symmetries
of these flows, which are restrictions to $u=\frac12$ of the linear
flows defined by \rref{CS+2}.
\endpf
A further outcome the previous construction is to provide a Lax representation
of the (extended) local CH hierarchy as a suitable flow in the space of
pseudodifferential operators. We will basically follow a construction
presented in  \cite{CFMPBrasil} for the KdV--KP case.

The Riccati constraint (\ref{ourcon}) can be read as the requirement
that the function $\psi=\exp{(\int h dx)}$ be an eigenfunction of
the operator $L=\frac{1}{m}\partial_x^2 - \frac{2u}{m}$ with
eigenvalue $z^2$.  Also,  the
equations of motion imply $\partial_s J^{(r)}=\partial_r J^{(s)}$
and $\partial_s h=\partial_x J^{(s)}$. Therefore, from the
compatibility of equations $L \psi = z^2 \psi$ and $\partial_s \psi
= J^{(s)} \psi$, we get
\begin{equation}
\label{interim}
\partial_{2s+1}L=\left[J^{(2s+1)},L \right] \qquad s \geq 1,
\end{equation}
while times and currents with even label $2s$ are trivial, as
implied by
the constraint (\ref{ourcon}).
In order to obtain an operatorial version of the equations
of motion we relate the currents $J^{(s)}$ with $L$.\\
First of all we need the following technical
 \begin{lem}
Under the constraint (\ref{ourcon}) it holds $J^{(s)}=\Pi_{J_+}(z^s)$.
\end{lem}
{\bf Proof }
The space $J_+=\Pi_{J_+}(J)$ is, by definition, the linear span of $J^{(i)}$.
Therefore there is a unique way to
write the element $\Pi_{J_+}(z^s)$ by means of the currents $J^{(s)}$.
Since the leading term of $J^{(s)}$ is
exactly $z^s$, the assertion is true.
\endpf
Because of Lemma \ref{reclem},
$J_+$ is also the linear span of the $\{(\partial_x+h)^i z^2\}_{i \geq 0}$. Moreover, extending by recursion
the definition of $(\partial_x+h)^i z^2$ to negative powers, the set  $\{(\partial_x+h)^i z^2\}_{i \in \mathbb{Z}}$
is a basis of all the space $J$.
The map
\begin{eqnarray*}
&&\phi:J \to \Psi DO \\
&& \phantom{\phi:} (\partial_x+h)^{i} z^2 \to \partial_x^i \cdot L
\end{eqnarray*}
by means of the basis $(\partial_x+h)^n z^2$
with $n \in \mathbb{Z}$ of the space $J$, gives the operatorial
action of an element $J$ on $\psi$
\begin{prop}Under the constraint (\ref{ourcon}) it holds
$$
J^{(s)} \psi = \left( L^{s/2-1} \right)_+ L \psi
$$
\end{prop}
{\bf Proof }
The map $\phi$ intertwines between $\Pi_{J_+}$ and the
operator  $(\ \cdot \ L^{-1})_+ L$ on the $\Psi$DO space
where $(\ \cdot\ )_+$ is the standard projection on
the differential part of a $\Psi$DO operator.
This property can be easily proved remarking that it holds for any element $(\partial_x+h)^i z^2$ of the $J$ basis.
Therefore
\begin{eqnarray*}
J^{(s)} \psi = \phi(J^{(s)}) \psi=\phi(\Pi_{J_+}(z^s)) \psi=(L^{s/2-1})_+ L \psi.
\end{eqnarray*}
\endpf
The equations (\ref{interim}) become then
\begin{equation}
\partial_{2s+1} L = \left[(L^{s-1/2})_+L , L  \right].
\end{equation}
We finally notice that in the CH case that is, under the constraint
$u=\dsl{\frac12}$, the Lax operator is
$L=\dsl{\frac{1}{m}\partial_x^2 - \frac{1}{m}}$; therefore
$(L^{1/2})_+=m^{-1/2}\partial_x - \frac{1}{2} (m^{-1/2})_x$ and the
previous equation gives
$$
\partial_3 \frac{1}{m} = -2m^{-2}\left(\partial_x - \frac{1}{4}\partial_x^3 \right) m^{-1/2}
$$
which is equivalent to the local CH (\ref{locCH}).
%
\\ We end this Section noticing that the  integrability of the
system constructed starting from the Lax operator for local CH can be proven
also by means of a direct computation. Indeed it holds:
\begin{prop}
Let $DO_2$ be the space of second order differential operators of the form
$\lambda=a\partial^2+b\partial+c$, and let $()_+$ be the
projection operator from $\Psi DO$ to $DO$. The equations
$$
\partial_s \lambda =\left[(\lambda^{\frac{s}{2}})_+ \lambda , \lambda\right]
$$
define a family of commuting flows on $DO_2$, that is,
$\partial_r \partial_s \lambda=\partial_s \partial_r \lambda$.
\end{prop}
{\bf Proof}. We start expanding
\begin{eqnarray*}
\partial_r \partial_s \lambda &=&\partial_r \left[(\lambda^{\frac{s}{2}})_+ \lambda , \lambda\right]
=  \left[(\partial_r \lambda^{\frac{s}{2}})_+ \lambda , \lambda\right]+
 \left[( \lambda^{\frac{s}{2}})_+ \partial_r\lambda , \lambda \right]+
 \left[( \lambda^{\frac{s}{2}})_+ \lambda , \partial_r\lambda \right] \\
&=&  \left[ \left[(\lambda^{\frac{r}{2}})_+ \lambda , \lambda^{\frac{s}{2}}\right]_+ \lambda , \lambda\right]+
 \left[( \lambda^{\frac{s}{2}})_+ \left[(\lambda^{\frac{r}{2}})_+ \lambda , \lambda\right] , \lambda \right]+
 \left[( \lambda^{\frac{s}{2}})_+ \lambda , \left[(\lambda^{\frac{r}{2}})_+ \lambda , \lambda\right] \right],
\end{eqnarray*}
as well as  $\partial_s \partial_r \lambda$. Then the assertion follows using
standard techniques in the $\Psi DO$ approach to KP-type equations (see e.g. \cite{DJKM}), with the
crucial remarks that, since we are considering degree $2$ operators,
\[
\left( \left[ (\lambda^{\frac{r}{2}})_- \lambda , (\lambda^{\frac{s}{2}})_- \right] \right)_+
=\left( (\lambda^{\frac{r}{2}})_- \left[ (\lambda^{\frac{s}{2}})_-, \lambda \right]\right)_+ =0,
\]
because the degrees of the operators appearing in these expression
is less than zero.
\endpf

\smallskip {\bf Acknowledgements } The authors would like to thank Marco
Pedroni, Boris Dubrovin, Andrew Hone, Paolo Lorenzoni, and Franco
Magri for useful discussions and remarks. G.O also benefited from
discussions with Paolo Casati and Boris Konopelchenko, and thanks
the University of Milano Bicocca for the kind hospitality, as well
as the organizer of the NEEDS 2007 Conference. This paper was
partially supported by the European Community through the FP6 Marie
Curie RTN {\em ENIGMA} (Contract number MRTN-CT-2004-5652), by the
European Science Foundation  project {\em MISGAM}, and by the
Italian MIUR Cofin2006 project ``Geometrical methods in the theory
of nonlinear waves and applications''.

\label{lastpage}

\begin{thebibliography}{99}

\bibitem{CamHol} R. Camassa, D.D. Holm,
\emph{An integrable shallow water equation with peaked solitons.}
Phys. Rev. Lett.  {\bf 71}  (1993),  no. 11, 1661--1664.

\bibitem{CFMPBrasil}  P. Casati, G. Falqui, F. Magri, M. Pedroni,
\emph{Soliton equations, bi--Hamiltonian manifolds and integrability}
21$\sp {\rm o}$ Col\'oquio Brasileiro de Matem\'atica. (21st Brazilian Mathematics Colloquium)
 Instituto de Matem�tica Pura e Aplicada (IMPA), Rio de Janeiro, 1997. iv+42 pp. ISBN: 85-244-0128-1

\bibitem{CFMP6}P. Casati, G. Falqui, F. Magri, M. Pedroni, {\em A Note on Fractional
KdV Hierarchies.} J. Math. Phys. {\bf 38}, 4606-4628, (1997).

\bibitem{CLOP} P. Casati , P. Lorenzoni , G. Ortenzi , M. Pedroni ,
\emph{ On the local and nonlocal Camassa-Holm hierarchies},
 J. Math. Phys.  46  (2005),  no. 4, 042704, 8 pp.

\bibitem{C}  A. Constantin, \emph{On the inverse spectral problem for the Camassa-Holm equation}
  J. Funct. Anal.  155  (1998),  no. 2, 352--363.

\bibitem{CmK} A. Constantin, H. P. McKean, \emph{ A shallow water equation on the circle}
Comm. Pure Appl. Math.  52  (1999),  no. 8, 949--982.

\bibitem{DJKM} E.Date, M.Jimbo, M.Kashiwara and T.Miwa,
\emph{Transformation groups for soliton equations},
in Non linear Integrable Systems, M.Jimbo and T.Miwa (eds.), World Scientific, Singapore 1983

\bibitem{DubZh} B.Dubrovin, Y. Zhang, \emph{Normal forms of hierarchies of integrable PDEs, Frobenius manifolds and
Gromov-Witten invariants} math.DG/0108160

\bibitem{FHO}, G. Falqui, A. Hone, G. Ortenzi, {work in progress}.

\bibitem{FF81}
B. Fuchssteiner, A.S. Fokas, {\em Symplectic structures,
their B\"acklund transformations and hereditary symmetries}.
Physica {\bf D},  4  (1981/82), 47--66.

\bibitem{FMP} G. Falqui, F. Magri, M. Pedroni,
\emph{Bihamiltonian geometry, Darboux coverings and
Linearization of the KP hierarchy.}
Commun. Math. Phys. {\bf 197} (1998), 303--324.

\bibitem{KM} B. Khesin, G. Misiolek, \emph{Euler equations on homogeneous spaces and Virasoro orbits.}
Adv. Math. {\bf 176} (2003), 116--144.

\bibitem{KO} B. Konopelchenko, W. Oevel \emph{An r-Matrix Approach to Nonstandard Classes of Integrable Equations}
 Publ. RIMS, Kyoto Univ. 29 (1993), pp. 581-666 

\bibitem{Len} J. Lenells, \emph{Conservation laws of the Camassa--Holm equation.}
J. Phys. A {\bf 38} (2005), 869--880.

\bibitem{OC} G. Ortenzi \emph{Some remarks on the KP system for the Camassa--Holm equation}, SIGMA 3 (2007) 047,
10 pages

\bibitem{PSZ} M.Pedroni, V. Sciacca, J. Zubelli, \emph{On the bi--Hamiltonian theory for the Harry--Dym equation}
Theor. Math. Phys. 133 (2002), 1583-1595

\bibitem{R} E.G. Reyes, \emph{Geometric integrability of the Camassa-Holm equation}, Lett. Math. Phys. 59 (2002), 117-131.

\bibitem{SW}G. Segal, G. Wilson, \emph{Loop Groups and equations of the KdV type}
Publ. Math. IHES, \textbf{61}, 5--65 (1985).

\end{thebibliography}
\end{document}